# Vacuum solutions of Einstein equations that depend on one coordinate

S. Parnovsky,
Astronomical Observatory, Taras Shevchenko National University of Kyiv, Kyiv, Ukraine

In the famous textbook written by Landau and Lifshitz all the vacuum metrics of the general theory of relativity are derived, which depend on one coordinate in the absence of a cosmological constant. Unfortunately, when considering these solutions, the authors missed some of the possible solutions discussed in this article. An exact solution is demonstrated, which is absent in the book by Landau and Lifshitz. It describes space-time with a gravitational wave of zero frequency. It is shown that there are no other solutions of this type than listed above and Minkowski's metrics. The list of vacuum metrics that depend on one coordinate is not complete without the solution provided in this paper.
Key words: general relativity, exact solutions

**Introduction**. The equations of the general theory of relativity (GRT) are a system of nonlinear equations, so finding their exact solutions is a difficult task. There are books entirely devoted to such solutions, for example [1]. The simplest solutions play a special role among them. These include, in particular, those whose metric depends on one variable. In this article, I consider the equations without the cosmological constant, as in the well-known textbook by Landau and Lifshitz [2] and use the same notation and choice of signs.

One can find all solutions whose metric depends on one variable for the case where there is no matter or other sources of the energy-momentum tensor in the right-hand side of Einstein's equations (a solution in a vacuum or a vacuum solution). This is described in §103 in [2] together with the case of a strong gravitational wave, when the coordinate on which the space-time metric depends is isotropic. In this paper, I demonstrate that this consideration is incomplete in the case when the coordinate is time-like or space-like and derive a vacuum metric that depends only on such a coordinate, which should be used to supplement the results of [2].

**Metrics considered in [2].** When considering the corresponding solutions, the authors of the textbook [2] are primarily interested in cosmological solutions, so they look for metrics that depend only on the time-like coordinate $t$. Therefore, the solution is sought in the synchronous system, the corresponding conditions (103.2-103.4) and the unnumbered equation following them are obtained, from which it is concluded that the determinant of the metric must be proportional to $t^2$. As a result, it is possible to obtain the well-known Kasner vacuum metric, which he derived back in 1921 [3]. It has the form

$$ds^2 = dt^2 - t^{2p_1} dx^2 - t^{2p_2} dy^2 - t^{2p_3} dz^2. \tag{1}$$

Coordinates $x, y, z$ are spatial. The constants denoted as $p_i$, $i = 1,2,3$ are so-called Kasner indices with two conditions

$$\sum_{i=1}^{3} p_i = p_1 + p_2 + p_3 = 1, \tag{2}$$

$$\sum_{i=1}^{3} p_i^2 = p_1^2 + p_2^2 + p_3^2 = 1. \tag{3}$$

Metric (1) corresponds to the vacuum solution for a homogeneous anisotropic cosmological model of Bianchi type I. By generalizing it, more complex solutions can be obtained, including the most general Belinsky-Khalatnikov-Lifshitz oscillatory solution near the cosmological singularity (see §113 in [2]). Thus, it is a very important metric underlying the consideration of anisotropic cosmological models.

But in the original work [3] Kazner obtained another metric, which is similar to (1) but depends on the spatial coordinate $x$ and has a naked time-like singularity at $x = 0$. It has a form

$$ds^2 = -dx^2 + x^{2p_1} dt^2 - x^{2p_2} dy^2 - x^{2p_3} dz^2. \tag{4}$$

Singularities at $t = 0$ in (1) and $x = 0$ in (4) are true ones except for cases $p_i = (1,0,0)$, $p_i = (0,1,0)$ or $p_i = (0,0,1)$. The curvature invariants diverge near these singularities.

Like other exact GR solutions, metric (4) requires a difficult analysis of the space-time it describes. This can be done on the basis of consideration of its geodesics [4]. The analysis is complicated by the infinite dimensions of the singularity. Therefore, its physical meaning is better investigated in the case of a similar singularity with a finite size, which is described by the so-called γ-metric. This was done in [5]. It is proved that at $p_1 < 0$ singularity is a point-like with negative mass, at $0 < p_1 < 2/3$ it is linear, and at $p_1 > 2/3$ it describes a space-time with a new type of singularity, which is impossible in a space with finite curvature. Such singularity was named paradoxical. Note that three singularities fundamentally different in their properties are described by one metric (4). This is a consequence of the nonlinearity of the GR equations, where the principle of superposition does not apply. In contrast to (1), properties (4) depend on the sequence of indices in the set, because the first of them is selected due to the signature of the metric. Generalizing (4), one can obtain an oscillatory solution near time-like singularities [6].

In addition to (4), there are two exact solutions of the GRT equations in vacuum, the metrics of which depend on one spatial variable. They are described in problems 1 and 2 after §104 in [2] and the corresponding space-times are studied in [7]. They correspond to the case when Kasner indices are complex or two of them are equal.

**A case not considered in [2].** But the given consideration is not exhaustive. This can be seen, for example, from the fact that there is no Minkowski metric among the obtained solutions. The problem is that the equation describing the dependence of the determinant of the metric on the coordinate *t* or *x* has another solution, too. It corresponds to the case when the determinant of the metric tensor is constant. Then it can be considered equal to -1 and instead of equations (103.7) in [2] or we have the following system

$$\gamma'_{\alpha\beta} = 2\lambda^{\delta}_{\alpha}\gamma_{\delta\beta}, \tag{5}$$

where the dash denotes the derivative with respect to the coordinate *x* on which the metric depends, the tensor $\gamma_{\alpha\beta}$ is a spatial part of the metric tensor and $\lambda^{\delta}_{\alpha}$ this is a constant matrix that satisfies two conditions

$$\lambda^{\alpha}_{\alpha} = \lambda^{\alpha}_{\beta}\lambda^{\beta}_{\alpha} = 0. \tag{6}$$

If its eigenvalues are real and they differ from each other, then the matrix can be reduced to diagonal form. From (6), we find that all eigenvalues in this case are equal to zero and the metric is a Minkovsky one.

If one eigenvalue is real and the other two are complex conjugates, then they are proportional to $\left(-2; 1+\sqrt{3}i; 1-\sqrt{3}i\right)$ due to (6). After the coordinate transformation, the metric of this vacuum solution takes the form

$$ds^2 = -dx^2 + e^{\alpha x}\left(\cos(\sqrt{3}\alpha x)(du^2 - dv^2) + 2\sin(\sqrt{3}\alpha x)dudv\right) - e^{-2\alpha x}dz^2. \tag{7}$$

Here α is an arbitrary constant. The curvature tensor is not zero, so it is not a Minkowski metric in a non-standard coordinate system. Curvature invariants are constant, there are no singularities. This is a strong gravitational wave of zero frequency. Coordinates *x* and *z* are spatial, and coordinates *u* and *v* change their type periodically with *x*.

Since the three real eigenvalues of the matrix $\lambda^{\delta}_{\alpha}$ coincide (they are equal to zero), by transforming the coordinates it can be brought to the form in which the elements on the diagonal and on one side of it are zero. From this canonical form, the conditions of preservation of the determinant and symmetry of the metric tensor, the metric can be reduced to

$$ds^2 = -dx^2 + 2dudv + Ax\,du^2 - dz^2, \tag{8}$$

where the constant *A* can be converted to 0, 1 or -1 by changing the coordinate scales. But the curvature tensor obtained by (8) is zero, that is, (8) is just the Minkowski metric in unusual coordinates.

**Conclusions**. I have completed the derivation of all possible vacuum metrics in the GRT without a cosmological constant, discussed in §103 in [2]. The solution (7), omitted in the book, was found. It describes a metric which depends on the spatial coordinate *x* only, the space-time has no singularities. This solution describes a gravitational wave of zero frequency. Unlike (1) and (4), it is not important in the study of space-time properties near singularities of arbitrary type. But (7) completes the list of metrics that have certain properties and it is not complete without it.